# Automation in Human-Machine Networks: How Increasing Machine Agency Affects Human Agency

Authors: Asbjørn Følstad[1], Vegard Engen[2], Ida Maria Haugstveit[1], Brian Pickering[2]

[1] SINTEF, Oslo, Norway

[2] IT Innovation, University of Southampton, Southampton, UK

**Abstract**

Efficient human-machine networks require productive interaction between human and machine actors. In this study, we address how a strengthening of machine agency, for example through increasing levels of automation, affect the human actors of the networks. Findings from case studies within air traffic management, crisis management, and crowd evacuation are presented, exemplifying how automation may strengthen the agency of human actors in the network through responsibility sharing and task allocation, and serve as a needed prerequisite of innovation and change.

## 1. Introduction

In the hyper-connected society, computing technology as part of human-machine networks (HMNs) (Eide et al., 2016) is ubiquitously available for supporting or conducting tasks previously dependent on human skill or expertise. In so doing, this increases machine agency (Engen et al., 2016) in the network. Machine actors typically do not replace human actors but rather take on dedicated tasks or support human actors in, e.g., information acquisition, analysis, or decision making, highlighting the need for productive interaction between human and machine actors.

We understand *agency* as capacity of the humans or machines in terms of what they can do and achieve in the network (Engen et al., 2016). Increasing levels of machine agency has raised concern regarding the need for human adaptation and flexibility (Lee & Moray, 1992), as well as its potential negative effects on human agency due to changes in situation awareness of human operators (Onnasch et al., 2014) and challenges to maintain human operator competency (Endsley and Jones, 2016).

The existing research is limited in that it typically concerns the effect of increasing machine agency in established systems with set goals and performance indicators. For example, in a meta-study on the effect of automation (Onnasch et al., 2014), the included studies typically concern whether automation can help reach existing goals in a more efficient manner, rather than how automation can motivate extending or changing the scope and goals of the HMN.

In this study, we address this limitation by conducting an integrated analysis of human and machine agency in the context of three case studies, to investigate how increased machine agency may strengthen innovation or change in HMNs. The findings from the three cases constitute the first steps towards an alternative perspective on the effects of automation; at the level of the entire HMN rather than limited to specific processes or existing goals. We thereby contribute to understanding the interplay between human and machine agency, and their implications for constructive interaction and innovation in HMNs.

## 2. Background

### 2.1 Agency in Human-Machine Networks

Eide et al. (2016) presented a typology of HMNs which is useful for understanding the potential effects of automation in such networks. *Human-machine networks* (HMN) are understood as "assemblages of humans and machines whose interaction have synergistic effects" (Tsvetkova et al., forthcoming), accentuating the potential for novel and improved outcomes enabled by such networks. The typology address the human and machine actors of the network, the relations between these actors, the extent of the network in terms of its size and geographical reach, and the network structure.

For the purpose of this paper, the typology dimensions concerning the agency of the network actors, as well as their relations are of particular interest. Drawing on the theory of *double dance of agency* (Rose and Jones, 2005), these dimensions serve to pinpoint the networked roles and interactions between humans and machines. The agency dimensions concern whether the actors typically engage in open and varied activities, aiming to influence others. The relation dimensions concern the level of intimacy and reciprocity between human actors, and reliance and dependency between human actors and machine actors.

### 2.2 Automation – changing the balance in human and machine agency

Automation may be seen as one step towards increasing the overlap between human and machine actors in HMNs, as human control, planning, and problem solving are replaced by machinery. In this context, much research has addressed the costs of automation. In the short term, increased automation may negatively affect situation awareness and performance under unexpected conditions or system failure; referred to as an out-of-the-loop syndrome (Onnasch et al., 2014). In the long term, automation may negatively affect human expertise or competency (Endsley and Jones, 2016). Mitigations to these challenges has been sought by, e.g., distinguishing between levels of automation (Endsley, 1999), introducing adaptive automation (Kaber and Endsley, 2004), and dynamic task delegation (Miller & Parasuraman, 2007).

This somewhat bleak picture of automation as an increase in machine agency at the cost of human agency is, however, far from complete. Wickens (Hancock et al., 2013) argue that the trade-offs typically associated with automation may not be inevitable. Furthermore, research on levels of automation (Onnasch et al., 2014) clearly indicate that automation limited to information acquisition and analysis typically strengthen routine performance while not affecting failure performance negatively.

### 2.3 Automation and impact on innovation and change

Automation is also seen as driver of innovation and change. Across nearly all sectors, machines are replacing humans for an increasing range of tasks, improving work effectiveness and efficiency. Such innovation and change is often seen as threatening for current employment patterns. However, in a recent canvassing of technology experts (Smith & Anderson, 2014), about half were optimistic concerning the impact of artificial intelligence and robotics on working life, suggesting that such technology will not replace more jobs than it creates and pave the way for innovation and growth. This extends Bainbridge's (1983) view that humans, due to automation, increasingly take on new tasks such as those related to system improvement. Likewise, Autor (2015) rhetorically asks why we still have so many jobs after decades of automation, and suggests that productivity-improvements in parts of a production process often serve to increase the value of the remaining parts, increasing the value of human skill and knowledge.

Hence, a fruitful perspective on automation may be to consider the interaction between increased machine agency on the one hand and the resulting effects on human agency on the other. A mere substitution on human actors with machine actors through increased automation would clearly reduce human agency. However, as new technology likely imply new opportunities for human actors, it may potentially strengthen rather than reduce human agency.

## 3. Research questions

On the basis of the background presented, we see the need to extend our knowledge base concerning how and why increased levels of machine agency in general, and automation in particular, may affect human agency in HMNs. Specifically, we need to investigate how and why increased levels of machine may produce synergetic effects with human agency in the network. For this purposes, we formulate the following research questions:

- How may the interrelation between human and machine actors be characterized in human-machine networks?
- How can automation be set up to strengthen synergy of machine and human actors of a human-machine network?
- How can automation be explicitly designed in order to strengthen capacities for innovation and change in the human-machine network?

## 4. Method

To address the research questions, three case studies (Yin, 2013) were conducted. Data collection and analysis were structured according to the HMN typology of Eide et al. (2016). In the following, the methods details for each case are briefly described.

Case 1 was conducted in connection to a research project concerning the development of a decision support system for air traffic management. The case involved semi-structured interviews and a subsequent workshop with four project representatives.

Case 2 concerns the use of a crisis management system to support information flow, shared situational awareness, and decision-making in public sector crisis management. The case involved interviews with six development representatives and six end-users, and a workshop with the development representatives.

Case 3 concerns the development and introduction of a decision support system for crowd management during emergency. The case involved two qualitative focus groups with a total of eight project representatives.

The interview and focus group data were transcribed and analysed through thematic analysis (Ezzy, 2002). Themes addressing the research question were identified, and the findings were validated with case study participants.

# 5. Findings

## 5.1 Case 1 – air traffic management

Case 1 is a research project aiming to develop automated decision support for operational air traffic management at airports. The stated aim of the project is to increase efficiency through automation while not compromising situation awareness and operator competence.

**Characterizing human and machine agency:** The interviewed researchers describe human agency as low in current operational air traffic management, largely determined by procedure and rules of thumb, due to the complex and safety critical character of the domain. The operators' main objective is to schedule and oversee the safe arrival and departure of the airplanes, and may have limited capacity and incitement for optimizing this schedule according to global performance criteria. Likewise, the machine agency of the current system is described as low. The operators have support systems for scheduling and monitoring of arrivals and departures, but while providing situational awareness these systems have low levels of intelligent decision support.

Low levels of intelligent decision support, combined with a highly procedural approach for scheduling and monitoring is reported to imply a challenge of decision making towards local optima, rather than a global optimum. Typically, scheduling a smooth sequencing of airplanes at the airport is implicitly prioritized above working towards maximum global efficiency in the network. As stated in one of the interviews:

> *In air traffic control, there are multiple operators working in sequence. And choices made early in the line may seem great there, but may force those later in the sequence to make suboptimal choices.*

**Strengthened synergy of human and machine actors:** The HMN typology was used to investigate how increased automation could support process transformation, e.g., enabling dynamic adjustment of human and machine agency and optimisation for global rather than local goals, whilst constrained by the very real need to preserve human actor oversight and governance. To strengthen the uptake of the decision support system, and keep up operator competence, the researchers report that the relation between operators and the decision support system should be one of collaboration. One vision for this collaboration is that the decision support system becomes a form of team member where humans and machines at times hold the same level of agency albeit at different levels of decision-making.

**Strengthen capacities for innovation and change:** The researchers report that rather than to increase machine agency at the expense of human agency, they would need to redefine the air traffic management process and allocate different responsibilities to human and machine actors. As stated in one of the interviews:

> *It is a gradual introduction of automation in the smaller decisions, so that the human can focus on larger, more critical, tactical, or strategic decisions.*

Specifically, an increase in machine agency, through increasing automated decision making at the local level and automatic negotiation of detail scheduling across locations, would allow operators to work more at a longer term tactical or strategic level across localities, allowing for exploring options with respect to a broader range of goals than today. Furthermore, such increase in machine agency is seen as instrumental for providing the human agency needed for change and innovation in air traffic management procedure.

## 5.2 Case 2 – public crisis management

Case 2 concerns a system for public crisis management that draws on information from multiple sources to facilitate dynamic coordination across the actors of a potential crisis situation. Information flow and obtaining shared situational awareness is considered crucial for decision-making.

**Characterizing human and machine agency**: The crisis management system and its users comprise a HMN where human agency is reported as high, as the tasks of human actors are highly varied and dependent on the organizational context. Machine agency is considered low, as the system is predefined and configured by human users. One system representative stated the following:

> *I would say that the human actors have a quite high degree of freedom in the system. All decisions, mostly […], are made by humans. What happens automatically in the system is almost entirely a result of the actions that people make.*

The great degree of freedom for users in configuring the system requires deep knowledge both of the activities and tasks the system needs to support and of how to configure the system accordingly. The interviewed users reported that, although people within public crisis management are professionals with great knowledge and experience in preparing for and handling crisis situations, many lack the technical knowledge needed to configure the system adequately in response to a particular organizational context.

**Strengthened synergy of human and machine actors**: The users reported a need for strengthening the synergy between human and machine actors, and suggested this could be accomplished by increasing machine agency in the network. For example, the system could support crisis management personnel by automatically retrieving relevant action plans, contingency plans, and check lists relevant for an incident. As expressed by one of the users:

> *We would like to automate the incident potential based on the action plans […] We want smart systems that can do some of the thinking for us, but that also gives room for improvisation, meaning that we can stop it [the system].*

The users also expressed that having the system gather, structure, and share information related to the course of the crisis, could potentially support crisis management personnel in making better decisions and obtaining situation awareness.

**Strengthen capacities for innovation and change**: The participants suggested that a higher degree of automation could streamline and make HMNs for crisis management more efficient. By assigning suitable tasks to the system, the crisis management personnel would be given greater leeway to perform activities of a more tactical or strategic nature, such as making decisions or other tasks requiring analyses based on the human actors' experience and knowledge.

## 5.3 Case 3 – crowd evacuation

Case 3 focuses on a decision support system for effective crowd evacuation during emergency situations at venues such as airports and sport stadiums. The system aggregates and makes available real-time sensor data in support of operational staff. The case study looked in particular at trust implications: operators were concerned about job security; evacuee agency needs careful management; and collaborative models need to respond to dynamic changes in the different actors.

**Characterizing human and machine agency:** The HMN manifests two distinct states: (a) during normal operation when a venue is being monitored, and (b) during an emergency. Both human and machine agency are low but increase significantly from the first to the second state. The interviewed participants report that there is little need for human actors to intervene when monitoring, with machines reviewing data only to identify possible issues and human actors merely checking. By contrast, during an emergency the agency of operational and possibly emergency staff must increase along with greater machine processing to facilitate safe evacuation. Agency increases differentially, though, with greater real-time processing and two-way sensor control as well as increasing human decision making. For safety reasons, the staff seek to control the agency of evacuees; and for ethical reasons, of machines.

**Strengthened synergy of human and machine actors:** Increasing automation may depend on the type of venue. Airports, for instance, tend to rely on more rigid emergency processes, whereas at the other extreme, sports stadia still depend largely on manual intervention by stewards and other staff. Despite these domain differences, participants reported that automatic identification of emergent processes could significantly enhance the effectiveness of the system:

> […] you could also do an emergent real time [process for] your evacuees, if you identify an emergent movement you could actually focus your attention as a civil protection agency and guide them then because everyone is following. That's not something you could identify at design time […]

Furthermore, as new evacuation routes for instance are calculated by the system, smart signage or smart use of lighting could allow this information to be presented to evacuees immediately, preferably including explicit and transparent reasons such as blocked corridors.

**Strengthen capacities for innovation and change:** Increasing the synergy between the different actors in the network like this also leads to innovation and changes in the procedures, paving the way for other benefits: evacuees themselves might become active participants in the network. Further, though operational staff may fear that in certain circumstances they may be replaced by technology, increasing machine agency and thereby providing more information allows them more efficient and effective decision making.

The emergent real-time processing discussed above, allows significant change and innovation in evacuation processes, which could increase safety levels. However, the participants reported that there were caveats here. On the one hand, HMN owners may maintain staffing levels whilst increasing safety through better decision support, or alternatively reduce staff and save costs without reducing current safety levels. One of the participants made a note of a concrete example demonstrating how the decisions for increasing automation could be motivated by, e.g., enabling staff to monitor a greater number of CCTV cameras from a financial perspective:

> […] if I want to reduce cost what I might want to do is enable each member of operational staff to monitor five times more cameras, for instance.

In either case, the trust relationship between the decision support system and operational staff is important, as the staff need to rely on the information for their decision making that they are ultimately responsible and accountable for.

# 6. Discussion

The three presented cases provide complementary insight into the research question. In this section, we discuss findings across the cases and relate this to the existing literature.

## 6.1 The synergy of human and machine agency in human-machine networks

From the cases we have presented above, there are four clear types or considerations in respect of synergistic interworking between human and machine actors within HMNs. These include collaboration and responsibility sharing; the allocation of tasks; machine exploitation; and trust. All three cases highlighted that increasing agency would promote opportunities for human and machine actors within the network to work together to achieve common goals. This would require careful distribution of responsibilities, as seen in cases 2 and 3. However, it is also about recognising that increased agency provides not only for offloading repetitive tasks to machine nodes (as seen in Case 2), but rather to allow both humans and machines to focus on what they do best, providing mutual support, as seen in Cases 1 and 3. This is an interesting outcome that warrants further investigation, especially as the increase in dependency on machines is considered to having increased risks should they fail (Onnash et al., 2013). The case studies indicate that this would benefit from allowing increased machine-to-machine connectivity (Case 2), not least because machine components would be better equipped to aggregate data from multiple sources (Cases 2 and 3) to provide more effective support to human decision makers while increasing reliability. Clearly, promoting synergy between human and machine actors opens up greater scope for effective collaboration, which is consistent with the existing literature exploring human and machine agency (Rose & Jones 2005; Jia et al., 2012; Engen et al., 2016). Although this requires some level of trust, it may well help to maintain and even increase trust as human actors begin to appreciate machines as particularly valuable contributors to the effectiveness of the whole HMN rather than as competitors.

## 6.2 Strengthening the synergy between machine and human actors through automation

Increasing levels of automation inevitably leads to an increase in machine agency given the definitions proposed by Engen et al. (2016). The question is how that, in turn, affects human agency. The three cases suggest that task automation may provide a number of benefits. In the first instance, it provides support to human actors and thus increases and strengthens their agency as shown in Cases 1 and 3. Decision support systems will typically aggregate information from multiple sources as well as provide some level of analytical summary. In so doing, this provides valuable input for human actors to evaluate their options and reach more informed decisions. Beyond this, as all three cases show, the appropriate allocation of tasks to machine actors within the HMN means that human actors can focus more effectively on what they do best: making reasoned choices of how to act and respond with a clear ethical and affective perspective given all the information provided. Automation correctly thought out will therefore support and enhance human agency, but also introduce a collaborative interplay on the basis of what different actor types do well. In the case of the machines, this does go beyond simply increasing the amount and speed in which data can be processed as seen in the above case studies and recent literature (Tsvetkova et al., 2017), being able to solve increasingly complex data analytics problems. This ultimately provides an opportunity for innovation and emergent behaviours as we will describe in the following section.

## 6.3 Strengthening innovation and change through automation

While much research on automation in human-machine contexts have considered the possible costs of automation and how to mitigate these (Endsley and Jones, 2016; Onnasch et al., 2014), parts of the literature also accentuate the role of automation for innovation and change (Autor, 2015; Smith & Anderson, 2014). Our case findings serve to expand on how automation can take this role. Strengthening machine agency through automation may allow human actors to shift activities from the level of detail data collection and local procedure to higher level tactical or strategic tasks, which is seen in both Case 1 and 3. These cases exemplify Autor's (2015) point that automation does not reduce the need for human work, but serves to change its content; from the procedural towards the knowledge-based.

However, innovation and change due to automation may not only concern changes in work practices. Case 2 serves to illustrate how lack of automated support in knowledge-based work, such as the local adaptation of crisis management, may compromise current work practices due to a lack in technical knowledge. Here, strengthened machine agency may serve not only to change work processes but to improve their quality. The experiences from Case 2 serves to complement earlier findings where automation often leads to performance reduction during system failure (e.g. Onnasch et al., 2013). Here, performance issues are not due to abrupt system failure, but rather argued to be a possible consequence of a lack in automatic configuration and adaptation of the system to the requirements of the local context.

### 6.4 Limitations and future work

The present study is based on a small number of case studies that have similar characteristics; all three studies address domains concerning decision support. The findings related to the research questions need to be further validated. In future studies, attention should be put on exploring the proposed research questions within other types of HMNs, such as social media networks, knowledge creation networks, and sharing economy networks.

Furthermore, several questions reminds and need to be addressed in future research, e.g.: When and why can automation lead to strengthened synergy between human and machine actors of a human-machine network? When does the costs of strengthened machine agency outweigh the benefits? What are the contextual requirements and constraints? Answering these and similar questions will be able to provide a better understanding of how automation influences human agency and under which conditions.

Despite the limitations of this study, we argue that it serves as an initial step towards increased understanding of the interaction between human and machine agency, and has the potential to motivate future research on how to design for such interaction in HMNs.